\documentclass[9pt]{sig-alternate}
\pdfoutput=1 
\usepackage{epsfig,endnotes}
\usepackage{ifpdf}
\usepackage{cite}
\usepackage{multirow}
\usepackage{times}
\usepackage{url}
\usepackage{algorithm}
\usepackage{algorithmic}
\usepackage{paralist}
\usepackage{etoolbox}
\makeatletter
\patchcmd{\maketitle}{\@copyrightspace}{}{}{}
\makeatother
\newcommand{\subparagraph}{}
\usepackage{titlesec}
\usepackage[it]{caption}
\newcommand{\Chrome}{{Chrome}}
\newcommand{\ChromeMarket}{{Chrome Web Store}}
\newcommand{\Android}{{Android}}
\newcommand{\AndroidMarket}{{Google Play Store}}

\newcommand{\paragraphbe}[1]{\vspace{.75ex}\noindent{\bf \em #1} }

\begin{document}
%

\title{Apples and Oranges:\\ Detecting Least-Privilege Violators\\ with Peer Group Analysis}
\numberofauthors{3}

\author{
\alignauthor Suman Jana \\
       \affaddr{Columbia University}\\
\alignauthor \'{U}lfar Erlingsson\\
       \affaddr{Google}\\
\alignauthor Iulia Ion\\
       \affaddr{Google}\\
}

\maketitle

\begin{abstract}
Clustering software into \emph{peer groups} based on its apparent functionality allows for simple, intuitive 
categorization of software that can, in particular, help identify which software uses comparatively more privilege than  
is necessary to implement its functionality. Such relative comparison can improve the security of a software ecosystem
in a number of ways. For example, it can allow market operators to incentivize software developers
to adhere to the principle of least privilege, e.g., by encouraging users to use alternative, less-privileged applications 
for any desired functionality.  This paper introduces \emph{software peer group analysis}, a novel technique to identify 
least privilege violation and rank software based on the severity of the violation. We show that peer group analysis is 
an effective tool for detecting and estimating the severity of least privilege violation. It provides intuitive, meaningful results,
even across different definitions of peer groups and security-relevant privileges. Our evaluation is based on empirically 
applying our analysis to over a million software items, in two different online software markets, and on a validation of our 
assumptions in a medium-scale user study.

%
%
\end{abstract}

\section{Introduction}
\label{intro}
Modern software platforms like \Chrome{} or \Android{} often provide mechanisms 
for restricting an application's ability to perform security-sensitive activity. This is usually done 
through a \emph{permission system} where the application must request and receive 
the necessary privilege to perform any security-sensitive operation. An application can receive its
desired privileges by requesting the appropriate permissions either statically, 
during installation, or dynamically, at runtime, depending on the platform. 
The requested permissions are granted to the application only if the system and 
the user approve the request.

Such permission systems are designed to encourage software developers to follow the 
principle of least privilege~\cite{saltzer1975protection}\textemdash design software to work with the minimal set 
of privileges needed for performing its functionality. Having application developers 
follow this principle helps reduce potential damage to users, by ensuring that 
applications can only abuse the subset of privileges they have received,
even if they get compromised, or turn malicious. In particular, in order to gain the 
full privileges to a system, an attacker would have to compromise multiple such 
applications. Thus, least-privileged applications make the attacker's job harder and  
provide increased security to the user. 

\subsection{The challenge of achieving least privilege}
In practice, creating least-privileged applications is a challenge:
 finding minimal set of privileges required to implement a given functionality, on 
a particular platform, can be an extremely difficult task. 
The most intuitive way of implementing some functionality 
may require significantly more privileges than the minimal set required
for alternate implementations. 
Moreover, the minimal privileges needed to implement some functionality are 
dependent on many factors, including the platform API, programming language, libraries, and 
the interactive user interface
exposed for the functionality. 

Consider the case of a \Chrome{} extension that 
translates the contents of a user-selected webpage from one language to another. One way of 
implementing such a functionality is to monitor all the webpages visited by the user and inject 
software scripts that run in the context of those webpages. This injected software can display a 
user interface element (e.g., a translate button) on each webpage to allow the user to trigger the translation of 
that webpage. Once the user clicks the button, the webpage is translated and the Document Object Model (DOM) 
of the webpage is updated with translated contents and are shown to the user. 
This implementation will require privileges to all the user's data on all webpages that the user visits,
because the injected script has full access to those pages. 

At first glance, it might seem that no translation functionality 
could be implemented without script injection, and the full privilege it implies. 
This, indeed, is a popular means of implementing such functionality.
However, on closer scrutiny, one can find implementations that require far less privilege. 
For example, the \Chrome{} platform allows extensions to
display a {\textit context menu} on any webpage visited by the user,
such that the menu is only displayed after some user interaction (e.g., a mouse click).
Such context menus, and the corresponding scripts, cannot access any contents of an underlying 
webpage, unless that content has been highlighted by the user before the menu was displayed;
such  highlighted content can be accessed from the menu's event-handler code. Therefore, a translation extension could operate 
via a menu, translate text that is highlighted by the user, and display the translation in an overlay window.
Such an implementation might serve many users' purposes equally well, but requires far less privilege than the previous approach.

As shown by this webpage translation example, implementing functionality with least privilege often requires significant 
effort from software developers, a deep understanding of the platform, and potentially a rethinking of user interactions.
Unfortunately, developers on existing platforms have few incentives to reduce the privileges that they request,
since only the user stands in the way of their software receiving any privilege they desire. As studies have shown,
most users will nearly always grant a privilege requested by software developer~\cite{felt2011effectiveness, felt2012android},
which is not surprising, given the intricacies involved in such decisions. The result can be a trend where developers 
that flout the principle of least privilege, reap significant benefits from the increased privileges of their software.
This trend can quickly become a self-reenforcing cycle that renders a platform's permission system irrelevant,
as all software ascends to having full privilege. Such a trend can be depressingly hard to combat,
since it is difficult to reward developers that strive to use least privilege:
even for the most technically sophisticated and security-conscious user,
it is extremely hard to evaluate how well a software application
has minimized its use of privileges, for its functionality. 

\subsection{Market-based approach to seeking least \\privilege}
\begin{figure}[!t]
\centering
\includegraphics[width=\columnwidth]{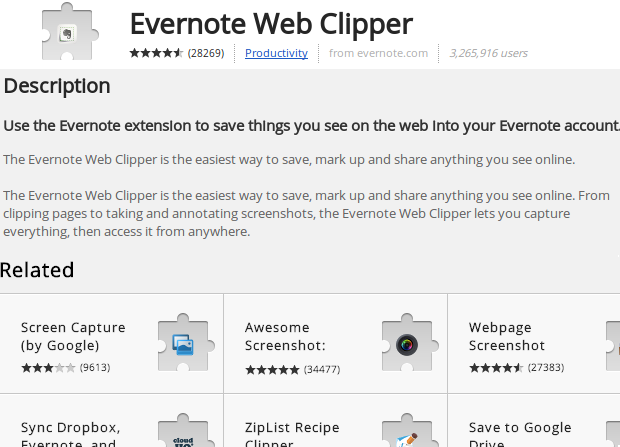}
\caption{The description and related items for the Evernote 
software in the Chrome Web Store.}
\label{fig:evernote_app_details}
\end{figure}

Unlike the developers and the users of the applications, the software markets hosting such applications 
have both the incentive and the means for automatically identifying the least-privilege violators. Online 
software markets have become a primary means 
by which end-user applications and other software are discovered, installed, updated, and managed. 
Such markets not only contain a great variety of software items (often in the millions), but also maintain 
a large amount of metadata about these items like descriptions and categorizations of applications, list 
of similar applications, etc. For example, Figure~\ref{fig:evernote_app_details} shows some of the 
metadata about Evernote Web Clipper software that appears in the Chrome Web Store.

In this paper, we demonstrate how software market owners can leverage market metadata to 
cluster hosted software into \emph{peer groups} based on functionality, analyze their privileges 
relative to the corresponding peer groups, and thus  identify over-privileged software in an automated 
fashion. A peer group consists of software that 
(for the most part) provide similar features and functionality, and for which the user can expect similar behavior.
For example, different text-editing applications are likely to belong in the same peer group,
whereas mobile-phone software that provides flashlight functionality belongs in another peer group.

Once peer groups have been defined, the common patterns of privileges 
can be identified for each peer group, and software ranked in terms of
how \emph{unexpected} its privileges are compared to its peer group.  Thus, the apples can 
be compared against the typical apple, and the oranges against the average orange.
Any outlier software who are significantly over-privileged can be isolated, and treated specially.  
As we show in this paper, such a ranking is helpful for multiple purposes like informing
the application developers about the unexpected privileges, 
helping users to pick lower-privileged software with the desired features, 
etc.


Software peer groups offer the most meaningful points of comparison
for software whose functionality is defined by a limited, well-defined set of related features.
Fortunately, most software in online software markets 
satisfies this property. In fact, for some markets, like the \ChromeMarket{},
the market operators explicitly prohibit software that bundles unrelated features,
as part of their fair use policy~\cite{policy_chrome}. 


We summarize our main contributions below.
\begin{itemize}
\item
We introduce the notion of peer group analysis in the context of detecting over-privileged 
applications. We evaluate our technique on two large datasets: one with 
$44,000$ \Chrome{} extensions from the \ChromeMarket{} 
and another with more than one million \Android{} applications from the \AndroidMarket{}. 
We find the analysis to be effective in identifying software violating the least privilege principle and identifying 
the unexpected privileges. 

We find that peer group analysis is a robust and generic technique:
irrespective of which piece(s) of market metadata we used to construct them,
our peer groups permitted effective estimates of spurious privileges,
as long as they correctly associated software with similar functionality.

\item
We demonstrate that the results of peer group analysis can be used for the following purposes: 
\begin{inparaenum}[\itshape a\upshape)] 
\item motivate the software developers to lower privileges of their applications by identifying 
over-privileged applications and their spurious privileges; and
\item provide users with lower-privileged alternatives for a desired feature by ranking all applications 
in the corresponding peer group according to their usage of spurious privileges. 
\end{inparaenum}

\item   
We  show that the idea of forming peer groups based on functionality is in sync 
with user expectations of software security privileges. Our user study confirm that 
users expect applications to have different privileges based on their functionality. 
We also confirm by analyzing some of the complaints filed by users of software 
in the \AndroidMarket{} that, in isolation, each user often has difficulty in correctly 
identifying the spurious privileges of different applications.


\end{itemize}

Our techniques have already been used in both \ChromeMarket{} and \AndroidMarket{} to 
rank, identify, and triage software with spurious privileges.

\section{Threat Model}
For this work, we assume that the application developers may be lazy 
or callous and violate the least privilege principle by developing  
over-privileged applications. However, we do not deal with actively 
malicious programmers who willingly steal user data or perform other 
abusive activities. Our goal is to create a safer ecosystem with 
increased incentives for non-malicious developers to create 
least-privileged applications.

\section{Detecting Over-Privileged Software with \\ Peer Group Analysis}
\label{pga}

Peer group analysis is a common technique in business economics to compare 
a corporation's financial performance (e.g., price-to-earnings ratio~\cite{pe_ratio}) 
against its peers with similar features like geographical location and industrial sector~\cite{peer_group_analysis}. 
In the context of anomaly detection, peer group analysis has been used by Bolton et al.~\cite{creditfraud} 
to detect credit card frauds. However, they tend to focus more on the temporal aspect of the peers' behavior. 
They formed peer groups by selecting the credit cards showing similar behaviors 
in the past and argued that their behavior should be similar in the future as well. 

By contrast, in this work, we apply peer group analysis in a way that is closer to its original usage 
in business economics. We identify peer groups of applications based on their 
functionality and compare their security privileges to get a better understanding 
of how security-sensitive a particular application's privileges are compared to those of its peers. 
Identifying the privileges commonly shared by the applications in the same peer group 
allows us to indirectly estimate the set of minimal privileges required for providing the common 
functionality of that peer group. Determining such mappings, without using peer group analysis, is 
extremely complicated and requires in-depth understanding of the platform API 
as well as the low-level details about the implementation of the functionality.

We define the following terms in the context of peer group analysis and use them for the rest of the paper. 
\begin{itemize}
\item
{\it Peer group.} Each peer group consists of a set of software applications
from which users expect to have similar functionality.

\item
{\it Unexpectedness.} For a given application, unexpectedness is the measure of
how over-privileged it is relative to its peers; this unexpectedness score is 
correlated with the severity of the least privilege violation in an application.

\item
{\it Unexpected privileges.} A security privilege used by an application is identified as 
unexpected if it is atypical relative to the security privileges of the application's peers.

\item
{\it Over-privileged  applications.} An application is assumed to be over-privileged if its 
unexpectedness value turns out to be greater than a certain threshold. The actual value 
of the threshold is a tunable parameter and may vary across different peer group analysis 
implementations.
\end{itemize}


Peer group analysis consists of two separate steps: identifying different peer groups based on 
functionality and finding over-privileged applications by detecting outliers in each such 
peer group based on their privileges. The list of over-privileged applications along with 
their unexpectedness scores can be used in several ways to make software platforms 
more secure, as described below.

\begin{itemize}
\item
{\it Providing incentives to developers for creating less-privileged applications.} 
Market owners can use peer group analysis to identify the applications who use 
less privileges than their peers. Such applications can be rewarded by ranking 
them higher in the search results and thus increasing their visibility. Similarly, 
the developers who repeatedly create severely over-privileged applications 
can be warned or suspended by the market owner.

Also, as part of computing the unexpected privileges, peer group analysis can 
estimate the minimal set of privileges used by applications in a peer group 
for implementing the common functionality. Sharing such information with a 
developer designing a new application that belongs to an existing peer group 
can be very useful as the list of minimal privileges provide the developer 
with a baseline to aim for. 

\item
{\it Helping users avoid over-privileged software.} As we show in Section~\ref{user_expectation}, 
security-conscious users try to detect spurious privileges of an application 
by associating its privileges with its functionality. Unfortunately, they often make mistakes 
due to the complexity of the process. Unexpectedness estimation using peer group analysis 
can automate this process and make it more robust. Sharing the unexpectedness scores 
with the users can help them find less privileged software for their desired functionality.

%

\end{itemize}

\section{Do Users Expect Security Privileges to be Tied to Software Functionality?}
\label{user_expectation}
In this section, we show that users expect software to use different privileges based on 
their functionality. 
To that end, we first analyze the abuse reports filed by users about 
the \AndroidMarket{} applications and show that users often complain about the 
security privileges of a software that they cannot relate to the functionality of the 
software. 
However, we also find that these
complaints often make
invalid assumptions
about the mapping between
security privileges and software functionality,
which are complicated and platform-dependent.
Next, we describe an Amazon Mechanical 
Turk (MTurk) study with $300$ participants whose results confirm that users indeed 
have different privilege expectations from applications providing different functionality.

\begin{figure*}[!t]
\centering
\includegraphics[width=\textwidth]{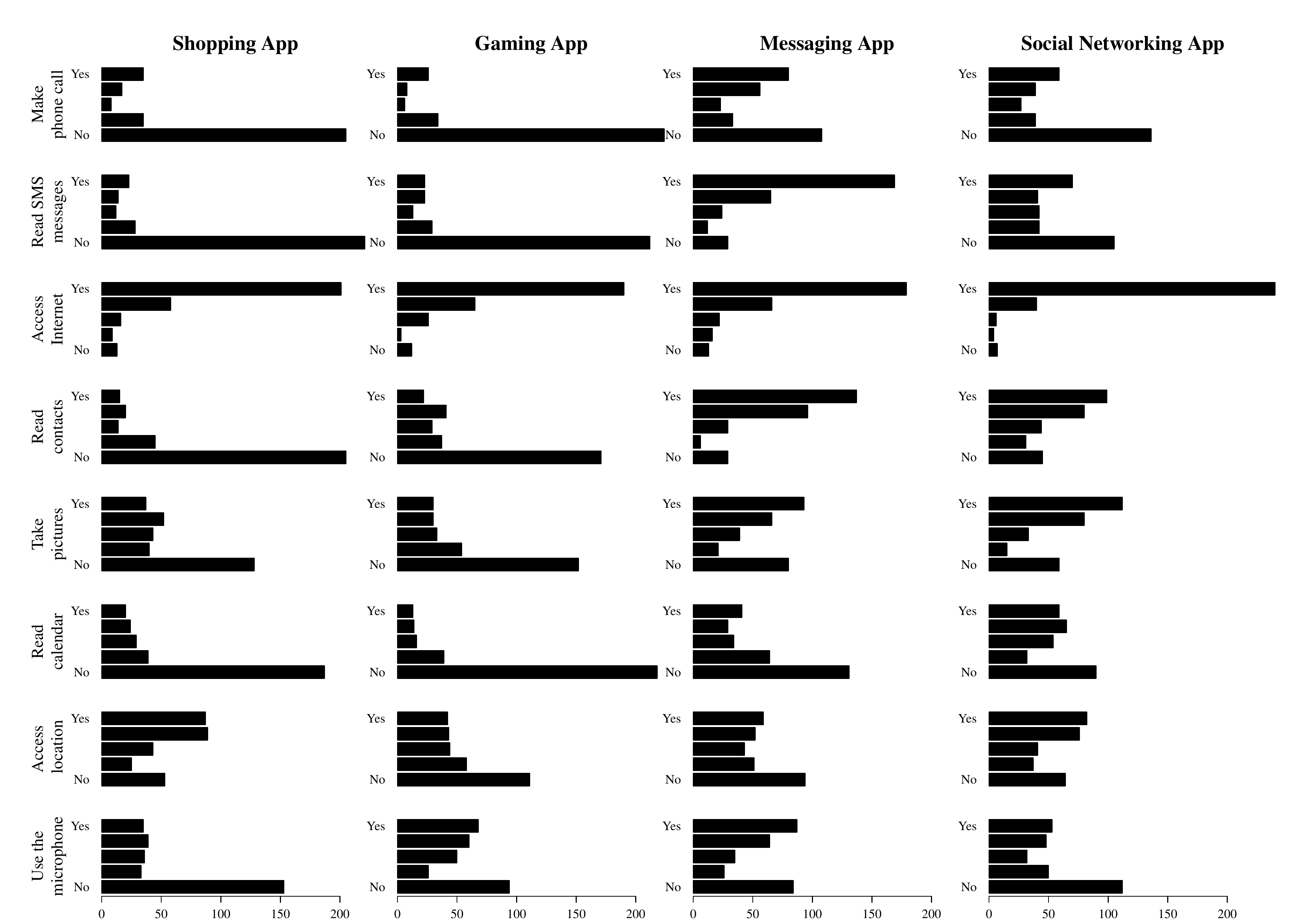}
\caption{User expectations of different security privileges (making a phone call, read SMS messages, access the internet, read the contacts on the phone, take pictures, read calendar entries, find out phone's location, and use the microphone) required by 
applications with different functionality (social networking, shopping, gaming, messaging).}
\label{fig:user_expectation}
\end{figure*}

\subsection{User reports of privilege abuse}
\label{abuse_reports}
We analyzed one month's submitted abuse reports, from the end of 2013,
where users of \AndroidMarket{} software
complain about abusive security permissions.
The dataset contained abuse reports for over $200$ different 
\Android{} applications, with between $1-40$ abuse reports per application. We find that, in general, 
users complained that applications require \emph{``too many permissions"} or that \emph{``permissions 
are far too invasive for what this application does.''} Many reports also explicitly stated 
several mismatches between the expected functionality and the requested privilege. For instance, one report said: 
\emph{``Please tell me WHY a banking app needs to take pictures or videos.''} Note that this user 
did not realize that this banking application needed the picture of a check for automatic check 
processing. Another report stated: \emph{``FLASHLIGHT app requires *network* access? Come on!''}. 
Similar to the previous case, the user did not understand that this application, like many other free ones, 
use network access for serving advertisements. 

These abuse reports show that users often try to detect unexpected privileges, depending on software functionality,
in line with previous findings by Jin et al~\cite{lin2012expectation}.
However, the reports also points out that the users need  significant assistance in order to correctly perform such 
analysis, e.g., by receiving a better basis for comparison from an automated system.

\subsection{ User study results }
We conducted an MTurk survey with $300$ participants to test if user expectations of the privileges required 
by specific applications depend on the application functionality. We had the survey reviewed 
by experts from our institution, and piloted with $20$ people on MTurk before launching it. Participants 
took on average $6$ minutes to complete the survey, and were compensated with $0.91$ US Dollar.

Previous studies showed that the MTurk population was more diverse than that found on a typical college 
campus and that using MTurk could result in high-quality data~\cite{Buhrmester2011amazon,Paolacci2011running}. 
However, as with any online survey, participants on MTurk may cheat by answering all questions quickly 
without reading thoroughly to collect the compensation. To prevent such spurious data affecting our results, 
we restricted our survey only to the respondents who had a task approval rate of $95\%$ or better and had completed 
at least $100$ tasks on MTurk. Furthermore, our survey had one trap question, which had one single, obviously right answer. 
We ensured that all participants answered that question correctly. Finally, one member of the research 
team reviewed all responses to the open-ended questions to ensure that responses were on topic.

\paragraphbe{Participant demographics.}
Participants in our MTurk survey were skewed slightly towards male and young: 68\% male, 32\% female; 32\%
were between the ages of 18-24, 41\% were 25-34, 18\% were 35-44, 5\% were 45-54, 3\% were 55-64, and 
1\% were 65 or over. As to their education, just over a third of the participants had Bachelor's 
degrees and another third had `some college'. The remaining third was spread over a broad range: from  
`some high school' (1\%) or a Doctoral degree (1\%) to having a Master's degree (8\%) or a regular high 
school diploma (9\%). Two thirds of the participants were employed full- or part-time or were self-employed 
(40\% full-time, 15\% part-time, and 13\% self-employed). 17\% of the participants were students, some of 
whom were also employed, and 16\% were unemployed or looking for work. Participants represented a broad 
range of occupations like driver, editor, photographer, library assistant etc. 


\paragraphbe{Results.}
Figure~\ref{fig:user_expectation} summarizes the results of our user study.
We asked the participants to rate on a 5 point Likert scale (`No', `Probably not', `Neutral', `Probably yes', and `Yes'), 
whether a \emph{social networking app}, a \emph{gaming app}, a \emph{messaging app}, and a \emph{shopping app} 
should be able to do a set of actions if installed on a cell phone. We intentionally did not use the worlds ``permissions'' 
or ``privileges," since previous research has found that not all users know what application permissions are~\cite{felt2012android}. 

Our results confirm that users expect applications with 
certain functionality to access specific resources on the phone, but not others. Participants almost unanimously expected a shopping and a gaming application to not be able to read SMS messages or 
read the contacts on the phone, while they thought a messaging application should be able to have both of these privileges. 
However, opinions about some privileges were mixed. For example, one participant said a shopping application should 
be able to find out the phone's location: \emph{``Absolutely needs location to find more relevant deals''} while another 
disagreed: \emph{``The app should not be able to access GPS. It should simply ask you the first time you install for your zip code''}.
Note that both agree that shopping applications can have access to some form of location information, but they differ 
in how fine-grained they think it should be. 

\section{Estimating Unexpectedness using Peer Group Analysis}
In this section, first we show how software market peer groups can be identified using several different 
sources of information that most markets already maintain about their hosted software.  Next, we 
enumerate different ways for approximating the security privileges of an application. 
Finally, we describe how one can compute an application's unexpectedness score using peer group 
analysis. 
 
\subsection{Identifying peer groups}
\label{identifying_peer_groups}
As peer groups are based on application functionality, in order to identify the peer groups, we must 
be able to the applications with similar functionality. Fortunately, most existing software markets 
maintain different sources of information about an application's functionality:  classification into pre-defined 
static categories, list of other related applications, textual descriptions, screenshots etc. The markets 
usually maintain such information to help users in finding out alternative applications providing similar functionality. 
We can simply leverage these existing sources of information to classify applications into different peer groups 
as described below. 

\paragraphbe{Classification provided by the developers.} Most markets including the \ChromeMarket{} and 
\AndroidMarket{} require the developers to classify their applications in one out of a fixed set of pre-defined 
categories while registering them with the markets. These categories are broad and each of them 
cover a large set of functionality. For example, \ChromeMarket{} supports different categories 
for applications like shopping, sports, news, blogging etc. We can create peer groups of applications using 
categories by simply putting all applications belonging to each category in one peer group. 



\begin{figure*}[!htbp]
\centering
\includegraphics[scale=0.35]{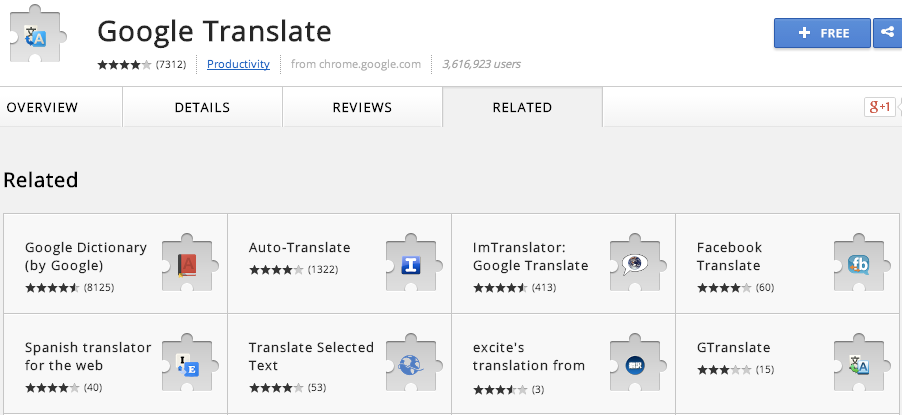}
\caption{The related software items for the Google Translate 
extension in the Chrome Web Store.} 
\label{fig:evernote_related}
\end{figure*}

\paragraphbe{Classification based on application recommendations.} Software markets usually maintain 
recommendation systems to help the users in finding new applications. Most modern recommendation systems 
output a list of related-items for the item an user is interested in. These related-items are usually computed 
based on collaborative filtering i.e. by extracting patterns from different users' behavior. For example, the \ChromeMarket{}
 displays a list of related applications when a user looks at the details of any particular application.  
Figure~\ref{fig:evernote_related} shows the list of related applications for the Google Translate extension
on the Chrome Web Store. One can clearly see that the related applications provide similar functionality i.e. translate text from one 
language to another. 

\begin{figure}
\centering
\includegraphics[width = 3.25in, height=2 in]{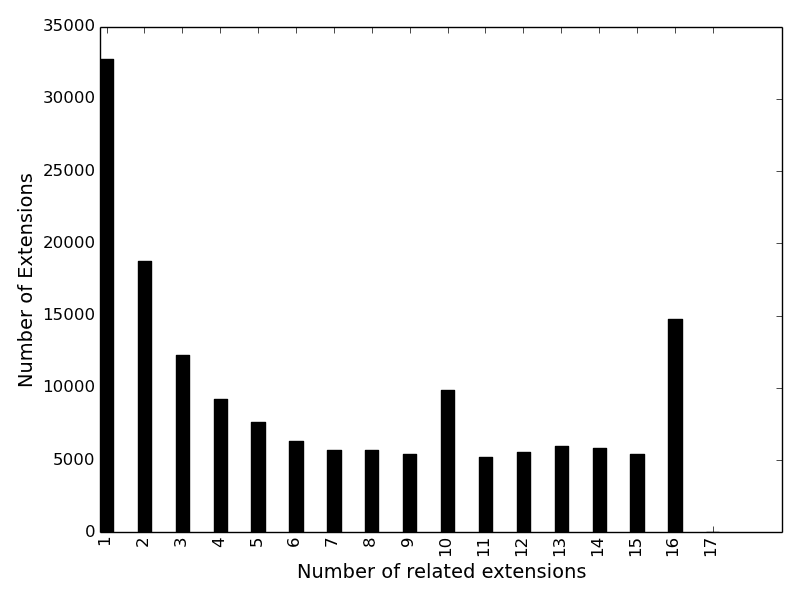}
\caption{Distribution of the number of related extensions extracted for each extension 
hosted in \ChromeMarket{}.}
\label{fig:related_items_crx}
\end{figure}

One simple way to create the peer group from the related application list is to put all related applications 
in one group. Figure~\ref{fig:related_items_crx} shows that this technique is very effective in creating 
small tightly-knit peer groups. 


\paragraphbe{Classification using other metadata.} Most software markets contain a large amount of 
unstructured metadata about each application like its textual description or user comments/reviews. These 
sources often contain a large amount of information about application functionality. Different automated 
classifiers can be designed to assign the applications to different peer groups based on these sources. 
For the rest of this subsection, we focus on identifying peer groups based on the textual descriptions of 
applications even though some of our techniques may also work on other sources like user reviews or 
comments.                          

In both the markets that we studied, all applications come with short textual descriptions about their functionality.
These textual descriptions are designed to make human users better understand application features.  

We explore two different techniques for designing classifiers that can use these textual descriptions 
to identify applications with similar functionality: one uses supervised learning algorithms to classify 
the applications into pre-defined categories and the other leverages unsupervised learning algorithms 
to cluster similar applications together. We describe both of these methods in detail below. 

\paragraphbe{Supervised classification.} Supervised classifiers can assign applications automatically to a set
of  pre-defined categories using features extracted from the textual descriptions of the applications.
However, such classifiers require a set of correctly labeled applications to train on, before they can be 
used for classification.  One way of creating such training data can be to simply use a small subset of the 
developer-provided categorizations from the market. Obviously, to ensure the quality of the training data,
 the categorizations should be manually checked.  

As a proof of concept, we built a Naive Bayes text classifier and used it to classify the extensions hosted 
in the \ChromeMarket{}. We gathered $15000$ extensions from the \ChromeMarket{} along 
with their textual descriptions and their developer-provided categorization in one of $19$ different categories. 
We further divided the dataset into two sets: a training set with $9000$ extensions and a test set 
with $6000$ extensions. Our Naive Bayes classifier is trained on the descriptions and categories of 
the extensions in the training set. Once trained, we use the classifier on the descriptions in the test 
set and evaluate the accuracy of the classifier by comparing the assigned categories with the 
developer-provided ones. In our tests, we found that our Naive Bayes classifier was able to correctly 
categorize almost $60\%$ of the extensions. While manually inspecting some of the applications that 
were assigned to different categories than their developer-provided ones, we found that our classifier 
actually categorized several extensions correctly that were mis-categorized by the developers. 
For example, the Google voice extension that can be used to make calls or send SMS was mis-categorized 
by the developer under the `blogging' category but the our classifier correctly assigned it to the 
`phone-and-sms' category.      


\begin{figure}
\centering
\includegraphics[scale=0.30]{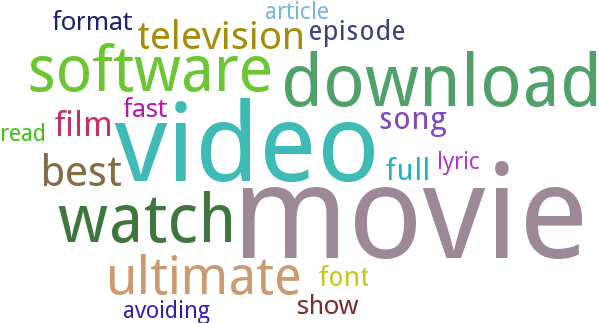}
\caption{Word cloud showing top 20 words of a topic detected by LDA that consists 
mostly of `movies' and `videos'. The font sizes of the words are proportional to their 
probability of selection under the topic.}
\label{fig:word_cloud_lda_movie}
\end{figure}

\paragraphbe{Unsupervised classification.} Unlike supervised techniques, the unsupervised ones do not need 
any training data and thus they can be used to detect applications with similar functionality without 
any manual effort. 

We built a prototype implementation using Latent Dirichlet Allocation (LDA)~\cite{blei2003latent} 
to both automatically find topics out of the textual descriptions and to estimate the likelihood of an 
application belonging to a topic. We classify the applications into different peer groups by creating 
one peer group for each such topic and including all applications whose probability of belonging to that topic 
is higher than a threshold. In order to evaluate our prototype, we collected the English textual descriptions 
of $44,000$ extensions from \ChromeMarket{}. We used standard natural language pre-processing 
techniques like removing the stop words and lemmatizing to clean up the descriptions. We also removed 
very short words with less than $3$ characters and words that do not appear in a standard English dictionary. 
Our implementation used the gensim library~\cite{gensim} for topic modeling to create a LDA model 
using the cleaned descriptions.  

Note that for creating meaningful peer groups, the topics found by LDA must map to real application 
functionality. In our experiments, we found that while some topics detected by LDA map to meaningful 
functionality, several topics do not relate to any particular functionality. For example, Figure~\ref{fig:word_cloud_lda_movie} 
show the word cloud representations of the topics found by LDA that map to `movies' category. However, as mentioned earlier, 
we also found that several topics generated by LDA did not relate to any particular 
functionality. For example, one such topic contained the following words 
in decreasing order of probability of their selection: `page', `click', `search', `link', `image',
`test', `button', `icon', `tab', and `site'.  Furthermore, when we looked at some of the 
developer-assigned categories of the extensions belonging to this topic, we found that $46\%$, $21\%$, $19\%$, 
$7\%$, and $7\%$ of them are classified by the developers under `productivity', `communication', 
`web development', `fun' and `search-tools' categories respectively. As the efficacy of peer group 
analysis depend on the ability to create peer groups of applications based on functionality, we chose 
not to use the LDA-based classifiers further in our experiments for the rest of the paper.



\subsection{\mbox{Estimating security privileges}}
Before we can compare an application against its peers, we must be able to enumerate all security-relevant 
behaviors of the application. A simple and effective way of circumscribing such behaviors is to create a list 
of the security privileges used by the application. Most modern software platforms including \Chrome{} and \Android{} 
use pre-defined permissions to restrict an application's privileges to access arbitrary resources. Therefore, 
in such systems, the set of permissions used by an application describes the privileges it used. However, finding 
the exact set of permissions that are actually used by an application is a hard problem. Below, we describe 
two different ways to approximate permissions used by applications.   

\paragraphbe{Requested Permissions.}  In most static permission-based systems like \Chrome{} or \Android{}, 
application developers must declare all the permissions that their applications need to operate correctly.
However, in most cases, the application writers tend to over-estimate the permissions used by 
their applications~\cite{felt2011android, au2012pscout}. 

\paragraphbe{Estimated Permissions.} Permissions used by an application can be 
estimated by first collecting all API calls that an application makes, then using a 
platform-specific mapping between the API calls and the permissions to enumerate 
the corresponding permissions~\cite{felt2011android, au2012pscout}. The API calls 
made by an application can either be estimated statically or dynamically. The static 
techniques often overestimate the API calls due to the presence of unused functions while 
the dynamic ones underestimate the API calls because of the lack of full coverage. 



\subsection{Estimating unexpectedness} 
\label{compute_unexpected}
For computing unexpectedness of an application's privileges relative to its peers, one can use 
standard machine learning techniques like One-Class Support Vector Machines (OC-SVM)~\cite{ocsvm}. 
However, one of the major drawbacks of these techniques is that the rationale behind their 
decisions is often hard to explain to a human. In order to be useful for our purposes, the reason 
why an application is considered  over-privileged must be easily explainable to both the developers 
and users. To achieve this goal, we chose to use a simple intuitive technique, 
as shown in Algorithm~\ref{alg:score}, for computing unexpectedness scores. Our technique 
is easy to understand and reason about. This is a necessary condition in our setting as the  
the unexpectedness scores are designed to be interpreted by the developers and users.
Moreover, our algorithm can easily be modified to find out the baseline privileges for each 
peer group. This information can be very useful to a developer designing a new application 
that belongs to an existing peer group. 

The basic intuition behind Algorithm~\ref{alg:score} is to first isolate the uncommon privileges for 
each peer group and then compute each application's unexpectedness score based on how many 
such uncommon privileges it uses. $W_p$ in Algorithm~\ref{alg:score} indicates the amount of 
weight we assign to each such uncommon privilege.  



\begin{algorithm}[!t]
\begin{algorithmic}
  \FORALL{application $a$ in market} 
    \STATE $unexpectedness_{a} \gets 0$
    \STATE $P_a \gets$ privileges used by $a$
    \STATE $g_a \gets$ peer group of $a$
    \FORALL{$p \in$ $P_a$}
      \STATE{$N_g \gets$ number of applications in $g_a$ }
      \STATE{$N_{gp} \gets$ number of applications in $g_a$ using $p$}
      \STATE{$X_{gp} \gets N_{gp} / N_{g}$}  
      \IF{$X_{gp} < relative\_frequency\_threshold$}
        \STATE $unexpectedness_{a} \gets$ $unexpectedness_{a} + W_p $
      \ENDIF 
    \ENDFOR
  \ENDFOR
\end{algorithmic}
\caption{Computing the unexpectedness value of an application $a$ with respect to peer group $g$.}
  \label{alg:score}
\end{algorithm}

\section{Experimental Setup}
\label{setup}
We evaluate our techniques on two different software markets: the \ChromeMarket{}
and the \AndroidMarket{}. We describe our experimental  setup for each of 
these markets in detail below.  

\paragraphbe{Extensions from \ChromeMarket{}.}
We collected a set of $44,000$ \Chrome{} extensions covering all the extensions that 
were published in the \ChromeMarket{} during early 2014. For each extension, we extracted
its developer-provided category, the list of related extensions, and the list of requested 
permissions from the \ChromeMarket{}. To extract the set of requested permissions, 
we first parse declared permissions from the extension manifests, removed any 
invalid permissions that the developers may have added by mistake. As specific host 
permissions are usually rarely repeated across applications, we also filtered out such 
permissions except $all\_urls$. 

   
\paragraphbe{\Android{} applications from \AndroidMarket{}.}
We gathered more than a million \Android{} applications covering all applications present in the \AndroidMarket{} 
during a specific day in the last six months. We extracted  the developer-provided category, the list of requested permissions 
from the manifest, and the application binary for each application. The application binaries were disassembled 
using the smali disassembler and analyzed to enumerate all API calls as well as the permissions required 
to successfully make those API calls. To estimate the permissions from API calls, we used the mapping of 
API calls to permissions provided by Au et al.\cite{au2012pscout}. For the rest of the paper, we refer to 
these sets of permissions as ``estimated permissions'' of an application. Note that neither the estimated 
permissions nor the requested permissions exactly represent the permissions used by an 
application as an application binary often contains code that do not get executed and developers 
often request permissions that are never used by their applications. Therefore, to get a better approximation 
of the actual used permissions, we also computed the intersection of requested and estimated permissions
for each application.

\paragraphbe{Implementation.} We implemented our techniques using two Map-Reduce tasks: one for computing 
relative frequency of privileges for each peer group and the other for computing unexpectedness values for 
each application and detecting over-privileged applications. We implemented Algorithm~\ref{alg:score} as part of 
the second MapReduce task. For all our tests, we set $W_p$ to $1$ for the privileges that we deemed security 
sensitive and to $0$ for the rest.  


\section{Evaluation} 
In this section, we first use our \Chrome{} extension and \Android{} application datasets to evaluate 
how the actual methods for peer group estimation and the values of different settings for peer group analysis affect 
the estimated unexpectedness scores. Next, we explore how useful the unexpectedness scores are for 
different purposes (e.g., helping developers adhere to the principle of least privilege, helping users avoid 
over-privileged applications) using the same datasets. 

%
\subsection{Effects of peer group parameters}
To estimate the effects of different settings of peer group analysis on its effectiveness, 
for each setting, we measure how many applications in our datasets have no unexpected 
security privileges relative to their respective peer groups. As these applications 
are deemed to be `normal' in their peer groups, we expect them to be the majority in our datasets. 
A low number of such applications will indicate that the peer groups are not well formed i.e. they contain 
applications with completely different functionality. 

\begin{figure}[!t]
\begin{center}
\includegraphics[scale=0.40]{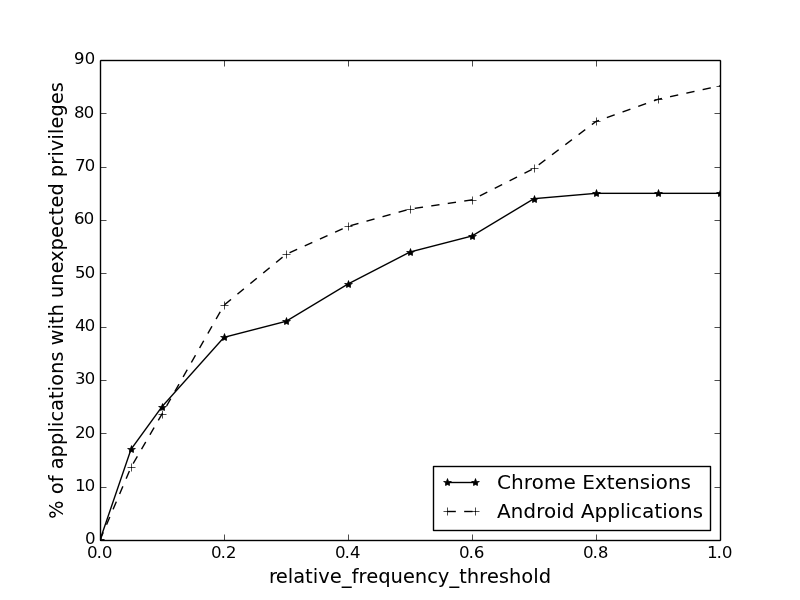}
\caption{Variability of the percentage of applications with at least one unexpected privileges, for different $relative\_frequency\_threshold$ choices.}
\label{fig:percentage_different_thresholds}
\end{center}
\end{figure}

\paragraphbe{Picking relative\_frequency\_threshold.} One of the main configuration parameters 
for our unexpectedness estimation algorithm (Algorithm~\ref{alg:score} in Section~\ref{compute_unexpected}) 
is $relative\_frequency\allowbreak\_threshold$. This parameter decides the minimum proportion of applications in a peer 
group that has to use a privilege in order to label that privilege as ``expected'' for that peer group. 
Figure~\ref{fig:percentage_different_thresholds} shows the variation of the percentage of applications 
that have at least one unexpected privilege with different values of the $relative\_frequency\_threshold$. 
For all of our tests, we set the relative frequency threshold to $0.10$ for \Chrome{} extensions 
and $0.05$ for \Android{} applications. 

\begin{figure}[!htbp]
\centering
\includegraphics[scale=0.40]{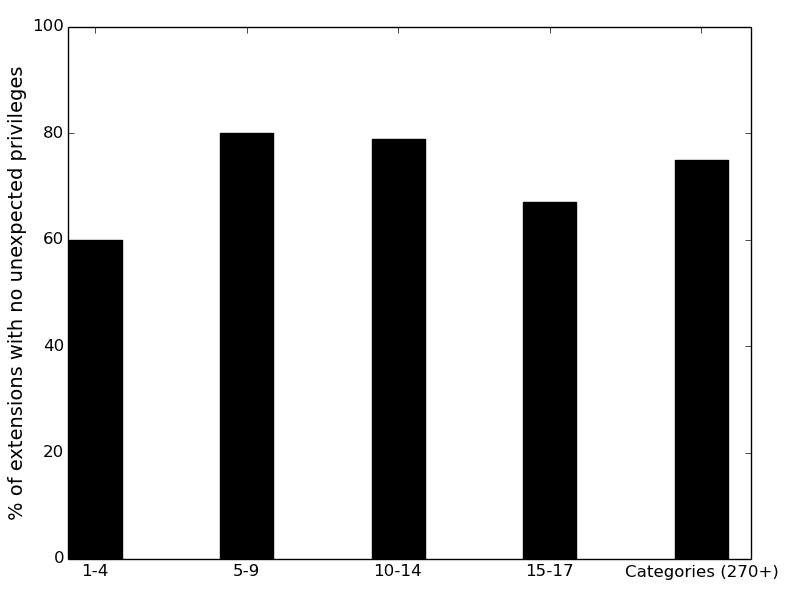}
\caption{Variability of the percentage of outlying extensions with unexpected privileges, for different peer group sizes.}
\label{fig:percentage_unexpected_peer_group_size}
\end{figure}

\paragraphbe{Peer group sizes.} Figure~\ref{fig:percentage_unexpected_peer_group_size} 
shows how the percentage of  \Chrome{} extensions with no unexpected privileges varies 
with different peer group sizes. For peer groups that contain only $1-4$ peers, there are only 
around $60\%$ of such extensions even for a low relative frequency threshold of $0.10$.  
This indicates that such small peer groups may not be very effective in estimating unexpectedness 
as they might mark a large number of applications as unexpected. However, with peer groups of 
size $10-14$, the percentage of extensions with no unexpected privileges rises to around $80\%$ 
for the same relative frequency threshold. Nonetheless, for peer groups of sizes larger than 
$14$, their percentage falls to around $60-70\%$.

\begin{figure}[!t]
\centering
\includegraphics[scale=0.40]{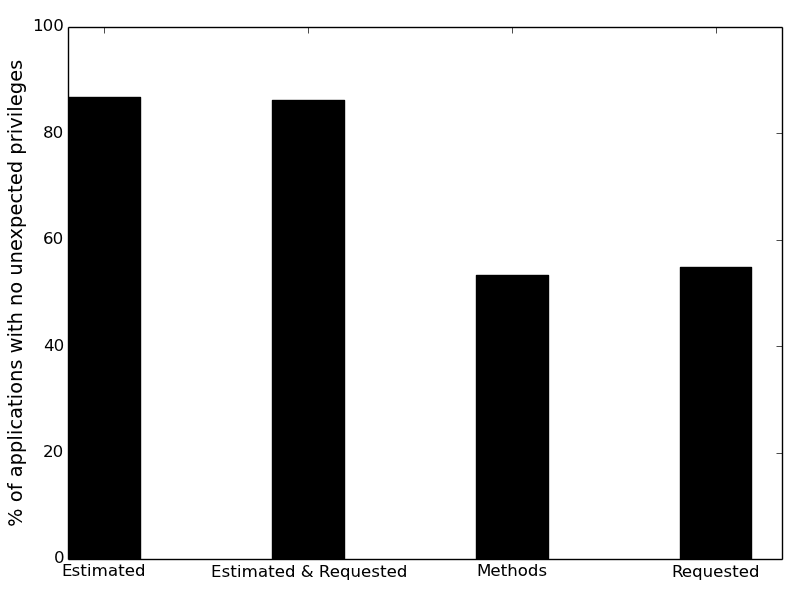}
\caption{Variability of the percentage of applications with no unexpected privileges, for different privilege types.}
\label{fig:percentage_unexpected_different_privileges}
\end{figure}

\paragraphbe{Different types of privilege estimation.} Figure~\ref{fig:percentage_unexpected_different_privileges} 
shows how the percentage of less-privileged applications vary with different types privilege estimation using the 
\Android{} dataset. We used four different ways for privilege estimation: requested privileges by the 
developers, method calls statically extracted from application binary,  statically estimated permissions from 
application binary, and intersection of statically estimated permissions and developer-requested permissions. 
We found that the statically estimated permissions and the intersection of the requested and estimated permissions 
yield the best results. 


\subsection{\mbox{Effectiveness of peer group analysis}}
\label{pga_effectiveness}
\paragraphbe{Provide incentives to developers for creating less-privileged applications.} 
Creating applications with lower privileges require significant amount of work and the existing system 
provides the developers no incentives to do it. Therefore, developers often tend to create over-privileged 
applications. However, we found that different developers tend to create differently over-privileged 
applications even when the applications have similar functionality. This is primarily caused 
by the differences in how the developers implement a particular functionality. We found wide 
variations in the unexpectedness scores of the applications from the same peer group.   
For example, Table~\ref{tbl:unexpected_chart} shows the unexpectedness scores and privileges of 
applications from two different peer groups in \ChromeMarket{}: one with five PDF readers and the other 
with eight tab managers. As we can see, different applications in the same peer group have widely different unexpectedness 
scores. For example, `docs-pdfpowerpoint-viewer' has a score of $0$ while  `pdf-viewer' has a score 
of $3$. Also, it can be seen form Table~\ref{tbl:unexpected_chart} that over-privileged applications 
in the same peer group like `pdf-viewer' and `\Chrome{}-office-viewer-beta' are over-privileged with 
very different privileges (pdf-viewer uses `WebNavigation' , `webRequest', and `webRequestBlocking' and
\Chrome{}-office-viewer-beta uses `clipboardWrite', `fileBrowserHandler', `fileSystem').

\begin{table*}[!htbp]
        \begin{center}
        \caption{Variation of unexpectedness across different peer groups.  
                 \label{tbl:unexpected_chart}
                }
                \begin{scriptsize}
                \begin{tabular}{|l|l|c|l|}
                        \hline
                        Peer group & App name & Score & Unexpected privileges \\ \hline
                        \multirow{5}{*}{pdf reader} & docs-pdfpowerpoint-viewer & 0 & - \\ \cline{2-4}
                                             & pdfescape-free-pdf-editor & 0 & - \\ \cline{2-4}
                                             & beeline-reader & 1 & webNavigation \\ \cline{2-4}
                                             & pdf-viewer & 3 & WebNavigation , webRequest, webRequestBlocking \\ \cline{2-4}
                                             & \Chrome{}-office-viewer-beta & 3 & clipboardWrite, fileBrowserHandler, fileSystem \\     
                                             \hline \hline                                             
                       \multirow{8}{*}{tab manager} & Tab Manager & 0 & - \\ \cline{2-4}
                                             & Project Tab Manager & 1 & bookmarks \\ \cline{2-4}
                                             & Awesome Tab Manager & 2 & bookmarks, unlimitedStorage \\ \cline{2-4}
                                             & TooManyTabs for \Chrome{} & 2 & bookmarks, unlimitedStorage \\ \cline{2-4}
                                             & Tabs Outliner & 3 & idle, notifications, storage \\ \cline{2-4}
                                             & Tabman Tabs Manager & 3 & history, topsites, webNavigation \\ \cline{2-4}
                                             & Fruumo Tab Manager & 3 & bookmarks, history, unlimitedStorage \\ \cline{2-4}
                                             & Session box - Tabs manager & 4 & clipboardWrite, cookies, management, unlimitedStorage \\ \hline
                \end{tabular}
                \end{scriptsize}
        \end{center}
\end{table*}

Such wide variation in the unexpectedness scores inside the same peer group also indicate 
that there is a large difference among the applications in terms of their adherence to 
the least privilege principle even when they are providing the same functionality. 
The unexpectedness score provides a metric to market owners for separating 
less-privileged applications from over-privileged ones. Market owners can leverage 
such information in different ways to encourage developers in creating less-privileged 
applications as mentioned below.
 
\paragraphbe{Rewarding applications with less privileges.} Applications with low unexpectedness
scores indicate that their developers have put more effort in following the principle of least privilege 
than other developers. To encourage such behavior, market owners can provide positive reinforcements like ranking such applications 
higher in the search results than their peers or providing the developers of such applications 
with special badges/points. Such positive reinforcements may encourage other developers 
to do a better job at creating less-privileged applications.

\paragraphbe{Penalizing callous developers.} An over-privileged application often 
indicates that its developer did not invest the required effort to lower its privileges. 
Among the over-privileged applications detected by peer group analysis, we found numerous 
such cases. For example, we found that the `pdf-viewer' uses the webRequest privilege to serve 
advertisements even when there are less-privileged ways of achieving the same goal.
As another example, consider the case of a Facebook game named `facebook mafia wars' that 
uses the privilege for installing or removing other extensions to scan and automatically 
remove any conflicting extensions. A less-privileged way of performing the same task would  
have been to simply ask the user to disable the conflicting extension. We also found that several 
developers repeatedly violate least-privilege in most of their applications irrespective of the 
functionality. Such developers can be identified easily with the unexpectedness scores of their 
applications and be warned or banned by market owners. 


\paragraphbe{Sharing information about peer groups with developers.}  In order to help developers 
create less-privileged applications, market owners can inform developers about the 
functionality and privileges used by the applications with low unexpectedness scores 
from different peer groups . Such information can be useful to the developers in several 
ways. First, a developer creating a new application can check the minimal of privileges 
used by applications in the corresponding peer group and treat these privileges 
as the baseline for the new application. Next, existing application developers 
may be more motivated to use less privileges if they know that a given functionality can be 
implemented using less privileges than they are using in their application. Consider three 
different Facebook notification extensions from \ChromeMarket{}: Facebook Chat Notification, Facebook notification icon, and 
Facebook Notifier; all of them provide similar functionality of notifying the user 
about some Facebook event (e.g., new chat messages, wall posts). However, the first 
one only requires the privilege to access the browser history, the second one requires access to 
the user's data from \url{facebook.com}, and the last one requires access to the user's data 
from all domains. If the developers of the second and third extensions 
know about the existence of the first one, they might be willing to emulate 
its implementation by redesigning their extensions to remain competitive.  

\begin{figure}[!htbp]
\centering
\includegraphics[scale=0.40]{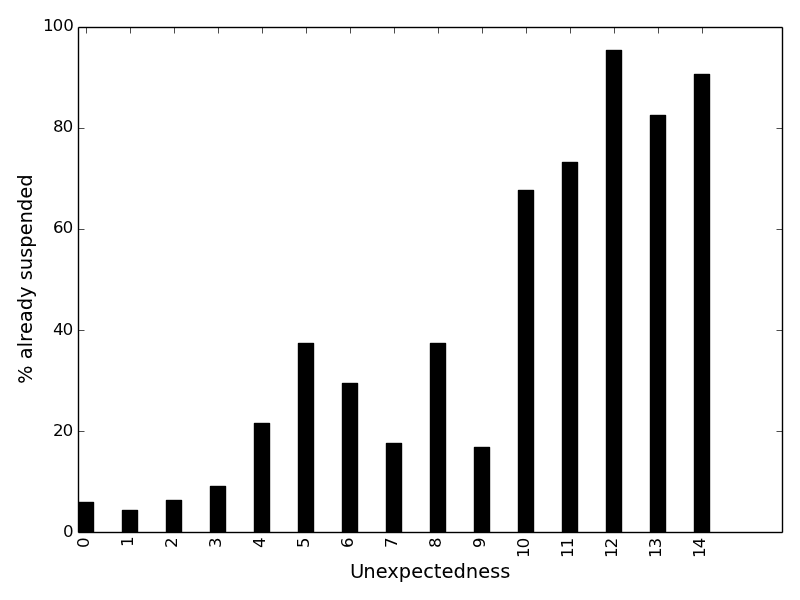}
\caption{Distribution of the percentages of suspended extensions with unexpectedness score in \ChromeMarket{}.}
\label{fig:banned_percentage_unexpectedness}
\end{figure}

\paragraphbe{Helping users to avoid over-privileged applications.} Over-privileged applications can 
cause significantly more damage to the user when compromised than their less-privileged counterparts. 
Moreover, we find  that over-privileged applications are more likely to be suspended by market owners 
due to policy violation, spamming, and other suspicious activities. Therefore, it is in the best interests 
of the users to try to avoid the over-privileged applications. 

\begin{figure}[!htbp]
\centering
\includegraphics[scale=0.40]{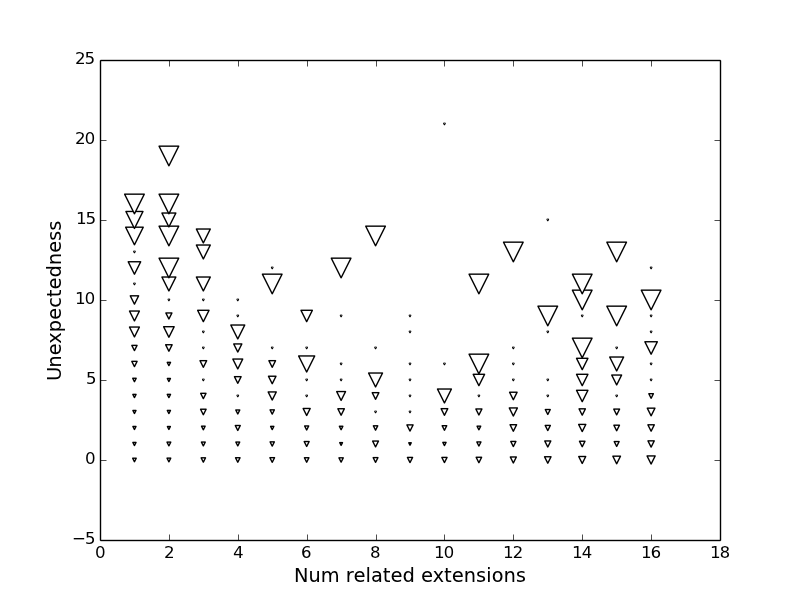}
\caption{Bubble plot showing variation of the percentage of suspended extensions in \ChromeMarket{} 
with unexpectedness score and  number of related items i.e. peer group size. The sizes of the 
triangles are proportional to the percentage of suspended extensions with a particular unexpectedness 
value and a particular peer group size.}
\label{fig:bubble_plot_banned}
\end{figure}

To evaluate the hypothesis that the over-privileged applications get suspended more often than th less 
privileged ones, we created a new dataset by picking $3828$ randomly chosen extensions from 
our existing \ChromeMarket{} dataset and augmenting them with  $3828$ suspended extensions. These 
extensions have been suspended by the market owner due to a variety of reasons including malicious 
activity, spamming, violating market policies etc.  We compute the unexpectedness values for all extensions 
in the dataset including both the suspended and live extensions. For each 
application with a unexpectedness score greater than zero, we compute the percentage of suspended applications 
for different scores. Figures~\ref{fig:banned_percentage_unexpectedness} and ~\ref{fig:bubble_plot_banned} 
show the results  from the experiments using the category-based peer groups and the peer groups based on 
related items respectively. Both these figures clearly demonstrate that, irrespective of the
technique used for creating peer groups, applications with high unexpectedness scores are very 
likely to be suspended. 

Unexpectedness score provides the users with a simple way to compare an application's privileges 
with respect to its peers (i.e., other applications providing similar functionality). A security-conscious 
user should pick applications with low unexpecteness scores from the peer group corresponding to the 
desired functionality. For example, if the user is looking for a PDF reader, she should probably pick 
the {\it pdf-viewer} application from the PDF reader peer group shown in Table~\ref{tbl:unexpected_chart}
(assuming, of course, it provides all the desired features).

\begin{figure*}[!tbp]
\centering
\includegraphics[scale=0.30]{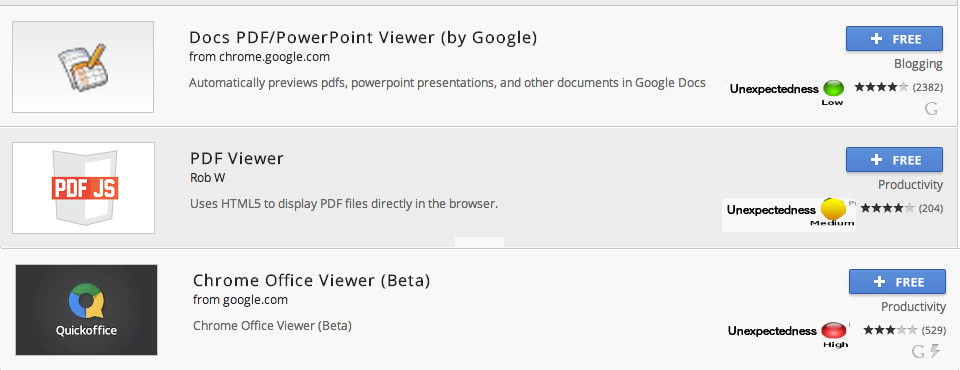}
\caption{A sample interface for displaying color-coded unexpectedness scores of each application 
present in the search results.} 
\label{fig:mockup_ui}
\end{figure*}

The market owners can facilitate this process by designing user interfaces that help users to 
easily find less privileged applications for a given functionality. For example, the search results 
returned to the user for a particular query to the application store can contain the unexpectedness 
scores of all applications present in the search results. The users should be able to 
sort the applications based on their unexpectedness score. Also, instead of showing 
the numeric unexpectedness scores to the user, the scores can be color coded 
according to their score (e.g., high, medium, and low scores can be represented by red, 
medium scores by red, yellow, green respectively as shown in Figure~\ref{fig:mockup_ui}).    

\section{Related Work}
\paragraphbe{Detecting malicious and risky software on Android.} 
Several prior works on the Android platform have focused on detecting malicious software using a variety 
of supervised or unsupervised machine learning algorithms~\cite{zhou2012hey, wu2012droidmat, arp2014drebin, 
zhou2012dissecting, rastogi, zhang2013vetting, felt2011malware}. 

One popular approach towards detecting potentially malicious Android applications is to identify risky combinations of Android 
permissions requested by an application and warn the users about them. A risk score is defined such that malicious applications 
tend to have higher risk scores than non-malicious ones. Such an approach has been argued to be more effective than the exiting 
system that shows warnings containing most of the requested permissions~\cite{sarma2012android, enck2009lightweight, grace2012riskranker}. 
However, identifying such combinations without considering application functionality often results in spurious warnings. To avoid such issues, 
machine learning algorithms have been used to analyze permission requests from different types of applications~\cite{
barrera2010methodology, ccs2012}. Peng et al.~\cite{ccs2012} introduced the idea of risk assessment of an Android 
application relative to other applications and evaluated different machine learning techniques for computing an application's 
risk score. They use the computed risk scores for detecting malware and protecting users from risky applications. 

However, all of these approaches assume that the application developers are malicious. By contrast, our approach deals with 
careless and lazy developers who are not actively malicious. Our goal is to make the developers, market owners, and users to 
work together to make more applications least-privileged.  In order to be helpful to the developers, the results of our approach 
must not only be explainable easily to the developers but also provide useful information for minimizing developer efforts spent 
in creating least-privileged applications (e.g., which privileges they should try to remove from an application).  As all the prior 
approaches simply focus on detecting malicious or risky software and helping users to avoid them, their techniques not suitable 
for our purposes. 

\paragraphbe{Detecting unused privileges.} At a higher level,  our goal to detect over-privileged applications is similar to 
that of Felt et al.~\cite{felt2011demyst} and Au et al.~\cite{au2012pscout}. However they used static analysis on the application 
code to find privileges that are requested by the application but are never used. By contrast, we focus on finding the privileges that are 
used by the application but are not necessary for the least-privileged implementation of the functionality. 

\paragraphbe{Mapping application privileges to their textual descriptions.} 
Pandita et al.~\cite{pandita2013whyper} and Qu et al.~\cite{qu2014autocog} used natural language processing techniques 
to identify if an application's description contain explanations about why they need certain permissions. They argued that 
sharing such information with the user can be very helpful in deciding whether the user wants to install the applications or not. 
Their approaches are complementary to ours and can be used to better explain to the users if an over-privileged application 
indeed needs an unexpected privilege for implementing a feature unique to its peer group. 

Gorla et al.~\cite{gorlachecking} built a tool called CHABADA that clusters Android applications with similar textual descriptions 
together using unsupervised machine learning techniques and identify outliers in each of the cluster according to their API usage 
for identifying malicious applications. While their goal is very different than ours, we evaluated the possibility of using textual 
descriptions of applications as a source of information for creating peer groups and found that such an approach, although somewhat 
promising, often put together applications with unrelated functionality in the same peer group due to lack of details about an 
application's functionality in its textual descriptions provided by the developer. We describe our findings in details in 
Section~\ref{identifying_peer_groups}.  Note that, for malware detection, it is immaterial whether all benign applications in the 
same cluster have similar functionality or not as long as all of their behaviors differ from those of the malicious applications 
in the cluster. Such an occurrence will ensure that the malicious applications will get detected as outliers even if the benign 
applications have different functionality. By contrast, for detecting least privilege violation, it is absolutely necessary for 
the applications in the same peer group to have similar functionality.

%

\paragraphbe{Detecting leakage of private information.} Detecting Android and iOS applications 
that leak user's private information has been another popular line of recent research~\cite{ yang2013appintent, 
gibler2012androidleaks, kim2012scandal, egele2011pios, encktaintdroid}. However, the detection of private data leakage is a 
related but different problem than the detection of least privilege violation as most such leaking applications 
are not violating the least privilege principle by accessing the private data i.e. they need access to the user's 
private data for providing their functionality.

\paragraphbe{User interfaces for Android permissions.} Several studies have found that users often have difficulty to 
comprehend the Android permissions shown to them during the installation process and tend to install applications 
irrespective of the permissions they request~\cite{felt2012android,felt2011android}. Peer group analysis can help 
users in estimating the risks of installing an application without understanding each of the permissions. 
Roesner et al.~\cite{roesner2012user} have proposed using access control gadgets provided by the operating system to allow users 
to grant permissions for user-controlled resources like camera in a non-intrusive and more intuitive manner.  
Such solutions can be used together with unexpectedness scores calculated by peer group analysis for helping 
users to control the sensitive privileges used by high-risk applications.

\section{Conclusion}
In this paper, we proposed and evaluated peer group analysis for effective and easy-to-understand detection 
of least privilege violation in applications hosted in online software markets. We showed that peer group 
analysis efficiently seeks out the over-privileged applications irrespective of the actual techniques used for 
forming the peer groups. We also showed that identification of over-privileged applications can help 
in improving the overall security of online markets by encouraging the developers to write less-privileged 
applications and the users to avoid over-privileged ones. Our technique has already been partially deployed 
in \ChromeMarket{} and \AndroidMarket{}. We hope that our work will encourage other software 
market owners to also adopt peer group analysis.\\

\noindent{\bf Acknowledgments.} 
We thank Karen Lees for her help in performing several experiments.

\bibliographystyle{abbrv}   
\bibliography{submission}  

\end{document}